\documentclass{cl2emult}

\usepackage{hyperref}
\usepackage{t1enc} % \k
\usepackage{amsmath}
\usepackage{latexsym}
\usepackage{theorem}
\usepackage{amssymb}
\usepackage{url}

\newcommand{\comment}[1]{}

\newcommand{\vsp}[1][3mm]{\vspace*{#1}}
\newcommand{\moins}{\setminus}
\newcommand{\vide}{\emptyset}

\newcommand{\yv}{\{y\to v\}}

\newcommand{\dom}{\mr{dom}}
\newcommand{\codom}{\mr{codom}}
\newcommand{\FV}{\mr{FV}}
\newcommand{\pos}{\mr{Pos}}

\renewcommand{\a}{\rightarrow}
\newcommand{\A}{\Rightarrow}

\renewcommand{\to}{\mapsto}

\newcommand{\I}[1]{[\![#1]\!]}

\newcommand{\sle}{\subseteq}

\newcommand{\tge}{\unrhd} % latexsym

\newcommand{\lex}{_\mr{lex}}

\renewcommand{\b}{\beta}

\renewcommand{\d}{\delta}

\newcommand{\vep}{\varepsilon}

\renewcommand{\t}{\theta}

\newcommand{\la}{\lambda}

\newcommand{\s}{\sigma}

\newcommand{\mc}{\mathcal}

\newcommand{\mr}{\mathrm}

\newcommand{\cC}{\mc{C}}
\newcommand{\cD}{\mc{D}}

\newcommand{\cF}{\mc{F}}

\newcommand{\cP}{\mc{P}}

\newcommand{\cR}{\mc{R}}

\newcommand{\cT}{\mc{T}}

\newcommand{\cV}{\mc{V}}

\newcommand{\cX}{\mc{X}}

\newcommand{\vl}{{\vec{l}}}

\newcommand{\vt}{{\vec{t}}}
\newcommand{\vu}{{\vec{u}}}

\newcommand{\vy}{{\vec{y}}}

\newcommand{\vT}{{\vec{T}}}

\newenvironment{rul}
  {$\begin{array}{rcl}}
  {\end{array}$}

\newenvironment{rew}[1][~~\a~~]
  {$\begin{array}{r@{#1}l}}
  {\end{array}$}
\newenvironment{rewc}[1][~~\a~~]
  {\begin{center}\begin{rew}[#1]}
  {\end{rew}\end{center}}

  {\end{array}$\end{center}}

\newcounter{counter}

{\theorembodyfont{\rmfamily} % theorem
  \newtheorem{dfn}[counter]{Definition}
  \newtheorem{lem}[counter]{Lemma}
  \newtheorem{thm}[counter]{Theorem}

}

\newenvironment{prf}{{\bf Proof.}}{}

\newenvironment{lstgeneric}[2]
  {\begin{list}{#1}{\topsep=.5mm\itemsep=.5mm\parsep=0mm%
    \itemindent=-3ex\labelsep=1ex\labelwidth=0ex #2}}
  {\end{list}}

\newenvironment{lst}[1]
  {\begin{lstgeneric}{#1}{\itemindent=-1ex}}
  {\end{lstgeneric}}

\newenvironment{enumi}[1]
  {\begin{lstgeneric}{}{\usecounter{enumi}\leftmargin=7mm%
    }}
  {\end{lstgeneric}}

\newcommand{\SN}{\mr{SN}}

\newcommand{\PCC}{\mr{PCC}}

\newcommand{\ch}{{\mr{C}}}

\newcommand{\acc}{\mr{Acc}}

\newcommand{\DP}{{\cD\cP}}
\newcommand{\cRi}{{\cR i}}

\newcommand{\horpo}{_{\mr{horpo}}}
\newcommand{\whorco}{_{\mr{whorco}}}

\newcommand{\ab}{\a_\b}
\newcommand{\ar}{\a_\cR}
\newcommand{\abr}{\a_{\b\cR}}
\newcommand{\ach}{\a_\ch}
\newcommand{\adp}{\a_\DP}
\newcommand{\adph}{\a_{\DP h}}
\newcommand{\ari}{\a_\cRi}

\newcommand{\abch}{\a_{\b\ch}}

\newcommand{\lx}{\la x}
\newcommand{\ly}{\la y}

\newcommand{\fap}{\mr{FAP}}

\newcommand{\level}{\mr{level}}

\begin{document}

%%%%%%%%%%%%%%%%%%%%%%%%%%%%%%%%%%%%%%%%%%%%%%%%%%%%%%%%%%%%%%%%%%%%%%%%%%%%%%
% title
%%%%%%%%%%%%%%%%%%%%%%%%%%%%%%%%%%%%%%%%%%%%%%%%%%%%%%%%%%%%%%%%%%%%%%%%%%%%%%

\title{Higher-order dependency pairs}

\author{Fr\'ed\'eric Blanqui}

\institute{LORIA\thanks{UMR 7503 CNRS-INPL-INRIA-Nancy2-UHP},
Campus Scientifique, BP 239, 54506 Vandoeuvre-l\`es-Nancy, France}

\maketitle

\begin{abstract}
Arts and Giesl proved that the termination of a first-order rewrite
system can be reduced to the study of its ``dependency pairs''. We
extend these results to rewrite systems on simply typed
$\lambda$-terms by using Tait's computability technique.
\end{abstract}

%%%%%%%%%%%%%%%%%%%%%%%%%%%%%%%%%%%%%%%%%%%%%%%%%%%%%%%%%%%%%%%%%%%%%%%%%%%%%%
% intro
%%%%%%%%%%%%%%%%%%%%%%%%%%%%%%%%%%%%%%%%%%%%%%%%%%%%%%%%%%%%%%%%%%%%%%%%%%%%%%

\section{Introduction}

Let $\cF$ be a set of function symbols, $\cX$ be a set of variables
and $\cR$ be a set of rewrite rules over the set $\cT(\cF,\cX)$ of
first-order terms. Let $\cD$ be the set of symbols occuring at the top
of a rule left hand-side and $\cC=\cF\moins\cD$. The set $\DP(\cR)$ of
{\em dependency pairs} of $\cR$ is the set of pairs $(l,t)$ such that
$l$ is the left hand-side of a rule $l\a r\in\cR$ and $t$ is a subterm
of $r$ headed by some symbol $f\in\cD$. The term $t$ represents a
potential recursive call. The chain relation is $\ach=\ari^*\adph$,
where $\ari^*$ is the reflexive and transitive closure of the
restriction of $\ar$ to non-top positions and $\adph$ is the
restriction of $\adp$ to top positions. Arts and Giesl prove in
\cite{arts00tcs} that $\ar$ is strongly normalizing (SN) (or terminating,
well-founded) iff the chain relation so is. Moreover, $\ach$ is
terminating if there is a weak reduction ordering $>$ such that
${\cR}\sle{\ge}$ and ${\DP(\cR)}\sle{>}$ (only dependency pairs need
to strictly decrease).

We would like to extend these results to higher-order rewriting. There
are several approaches to higher-order rewriting. In Higher-order
Rewrite Systems (HRSs) \cite{mayr98tcs}, terms and rules are simply
typed $\la$-terms in $\b$-normal $\eta$-long form, left hand-sides are
patterns \`a la Miller and matching is modulo $\b\eta$. An extension
of dependency pairs for HRSs is studied in
\cite{sakai01tis,sakai05tis}. In Combinatory Reduction Systems (CRSs)
\cite{klop93tcs}, terms are $\la$-terms, rules are $\la$-terms with
meta-variables, left hand-sides are patterns \`a la Miller and
matching uses $\alpha$-conversion and some variable occur-checks. The
relation between the two kinds of rewriting is studied in
\cite{oostrom93hoa}. It appears that the matching algorithms are
similar and that, in HRSs, one does more $\b$-reductions after having
applied the matching substitution. But, in both cases, $\b$-reduction
is used at the meta-level for normalizing right hand-sides after the
application of the matching substitution. So, a third more atomic
approach is to have no meta-level $\b$-reduction and add
$\b$-reduction at the object level. This is the approach that we
consider in this paper.

So, we assume given a set $\cR$ of rewrite rules made of simply typed
$\la$-terms and study the termination of $\ab\cup\ar$ when using
CRS-like matching. This clearly implies the termination of $\ar$ in
the corresponding CRS or HRS. Another advantage of this approach is
that we can rely on Tait's technique for proving termination
\cite{tait67jsl,blanqui05mscs}. This paper explores its use with
dependency pairs. This is in contrast with
\cite{sakai01tis,sakai05tis}.

In Tait's technique, to each type $T$, one associates a set $\I{T}$ of
terms of type $T$. Terms of $\I{T}$ are said {\em computable}. Before
giving some properties of computable terms, let us introduce a few
definitions. The sets $\pos^+(T)$ and $\pos^-(T)$ of {\em positive and
negative positions} in $T$ are defined as follows:

\begin{lst}{--}
\item $\pos^+(B)=\{\vep\}$ and $\pos^-(B)=\vide$ if $B$ is a base type,
\item $\pos^\d(T\A U)=1\cdot\pos^{-\d}(T)\cup 2\cdot\pos^\d(U)$.
\end{lst}

We use $\vT$ to denote a sequence of types $T_1,\ldots,T_n$ of length
$|\vT|=n$. The $i$-th argument of a function symbol $f:\vT\A B$ is
{\em accessible} if $B$ occurs only positively in $T_i$. Let $\acc(f)$
be the set of indexes of the accessible arguments of $f$. A base type
$B$ is {\em basic} if, for all $f:\vT\A B$ and $i\in\acc(f)$, $T_i$ is
a base type. After \cite{blanqui05mscs,blanqui06horpo}, given a
relation $R$, {\em computability wrt $R$} can be defined so that the
following properties are satisfied:

\begin{enumi}{}
\item\label{comp-sn}
A computable term is strongly normalizable wrt ${\ab}\cup{R}$.
\item\label{comp-basic}
A term of basic type is computable if it is $\SN$ wrt ${\ab}\cup{R}$.
%\item\label{comp-lambda}
%$\lx t$ is computable if, for all $u:T_x$ computable, $t_x^u$ is
%computable.
\item\label{comp-arrow}
A term $v^{T\A U}$ is computable if, for all $t^T$ computable, $vt$ is computable.
\item\label{comp-red}
If $t$ is computable then every reduct of $t$ is computable.
\item\label{comp-neutr}
A term $f\vt$ is computable if all its reducts wrt ${\ab}\cup{R}$ are computable.
%\item\label{comp-fun}
%$f$ is computable if, for all rule $f\vl\a r\in\cR$ and substitution
%$\s$, $r\s$ is computable whenever $\vl\s$ so are.
\item\label{comp-acc}
If $f\vt$ is computable then, for all $i\in\acc(f)$, $t_i$ is
computable.
\item\label{comp-no-fun}
If $t$ contains no $f\in\cD$ and $\s$ is computable, then $t\s$ is
computable.
\item\label{comp-all}
Every term is computable whenever every $f\in\cD$ is computable.
\end{enumi}

%%%%%%%%%%%%%%%%%%%%%%%%%%%%%%%%%%%%%%%%%%%%%%%%%%%%%%%%%%%%%%%%%%%%%%%%%%%%%%
% higher-order dependency pairs
%%%%%%%%%%%%%%%%%%%%%%%%%%%%%%%%%%%%%%%%%%%%%%%%%%%%%%%%%%%%%%%%%%%%%%%%%%%%%%

\section{Admissible rules}

An important property of the first-order case is that, given a term
$t$, a substitution $\s$ and a variable $x\in\cV(t)$, $x\s$ is
strongly normalizable whenever $t\s$ so is. This is not always true in
the higher-order case. So, we need to introduce some restrictions on
rules to keep this property.

%%%%%%%%%%%%%%%%%%%%%%%%%%%%%%%%%%%%%%%%%%%%%%%%%%%%%%%%%%%%%%%%%%%%%%%%%%%%%%
% rules for patterns
%%%%%%%%%%%%%%%%%%%%%%%%%%%%%%%%%%%%%%%%%%%%%%%%%%%%%%%%%%%%%%%%%%%%%%%%%%%%%%

\begin{figure}[ht]
\centering\caption{Pattern Computability Closure \cite{blanqui00rta}\label{fig-pat}}\vsp
\fbox{\begin{minipage}{11cm}
\centering\vsp
(arg)\quad $t_i\in\PCC(\vt)$\\[2mm]

(acc)\quad $\cfrac{g\vu\in\PCC(\vt)\quad i\in\acc(g)}
{u_i\in\PCC(\vt)}$\\[2mm]

(lam)\quad $\cfrac{\ly u\in\PCC(\vt)\quad y\notin\FV(\vt)}
{u\in\PCC(\vt)}$\\[2mm]

(app-left)\quad $\cfrac{uy\in\PCC(\vt)\quad
y\notin\FV(\vt)\cup\FV(u)}{u\in\PCC(\vt)}$\\[2mm]

(app-right)\quad $\cfrac{y^{U\A\vT\A U}u\in\PCC(\vt)\quad
y\notin\FV(\vt)\cup\FV(u)}{u\in\PCC(\vt)}$\vsp%\\[2mm]

%(subterm)\quad $\cfrac{
%u^B\tlt v\in\PCC(\vt)\quad B \mbox{ basic}\quad \FV(u)\sle\FV(v)}
%{u\in\PCC(\vt)}$\vsp
\end{minipage}}
\end{figure}

%%%%%%%%%%%%%%%%%%%%%%%%%%%%%%%%%%%%%%%%%%%%%%%%%%%%%%%%%%%%%%%%%%%%%%%%%%%%%%
% patterns
%%%%%%%%%%%%%%%%%%%%%%%%%%%%%%%%%%%%%%%%%%%%%%%%%%%%%%%%%%%%%%%%%%%%%%%%%%%%%%

\begin{dfn}[Admissible rules]
A rule $f\vl\a r$ is {\em admissible} if $\FV(r)\sle\PCC(\vl)$, where
$\PCC$ is defined in Figure \ref{fig-pat}.
\end{dfn}

The Pattern Computability Closure (PCC) is called accessibility in
\cite{blanqui00rta}. It includes most usual higher-order patterns
\cite{miller89elp}.

%%%%%%%%%%%%%%%%%%%%%%%%%%%%%%%%%%%%%%%%%%%%%%%%%%%%%%%%%%%%%%%%%%%%%%%%%%%%%%
% pattern computability
%%%%%%%%%%%%%%%%%%%%%%%%%%%%%%%%%%%%%%%%%%%%%%%%%%%%%%%%%%%%%%%%%%%%%%%%%%%%%%

\begin{lem}
\label{lem-comp-pat}
If $f\vl\a r$ is admissible, $\dom(\s)\sle\FV(\vl)$ and $\vl\s$ is
computable, then $\s|_{\FV(r)}$ is computable.
\end{lem}

\begin{prf}
We prove by induction that, for all $u\in\PCC(\vt)$ and computable
substitution $\t$ such that $\dom(\t)\sle\FV(u)\moins\FV(\vt)$,
$u\s\t$ is computable.

\begin{lst}{}
\item [(arg)] Since $\dom(\t)=\vide$, $l_i\s\t=l_i\s$ is computable
  by assumption.

\item [(acc)] By induction hypothesis, $g\vu\s$ is computable.
  Thus, by property (\ref{comp-acc}), $u_i\s$ is computable.

\item [(lam)] Let $\t'=\t|_{\dom(\t)\moins\{y\}}$.
  Wlog, we can assume that $y\notin\codom(\s\t)$. Hence, $(\ly
  u)\s\t'=\ly u\s\t'$. Now, since $\dom(\t)\sle\FV(u)\moins\FV(\vt)$,
  $\dom(\t')\sle\FV(\ly u)\moins\FV(\vt)$. Thus, by induction
  hypothesis, $\ly u\s\t'$ is computable. Since $y\t$ is computable,
  by (\ref{comp-arrow}), $(\ly u\s\t')y\t$ is computable and, by
  (\ref{comp-red}), $u\s\t'\{y\to y\t\}$ is computable. Finally, since
  $y\notin\dom(\s\t')\cup\codom(\s\t')$, $u\s\t'\{y\to y\t\}=u\s\t$.

\item [(app-left)] Let $v:T_y$ computable and $\t'=\t\cup\yv$.
  Since $\dom(\t)\sle\FV(u)\moins\FV(\vt)$ and $y\notin\FV(\vt)$,
  $\dom(\t')=\dom(\t)\cup\{y\}\sle \FV(uy)\moins\FV(\vt)$. Thus, by
  induction hypothesis, $(uy)\s\t'=u\s\t'v$ is computable. Since
  $y\notin\FV(u)$, $u\s\t'=u\s\t$. Thus, $u\s\t$ is computable.

\item [(app-right)] Let $v=\lx^U\la\vy^\vT x$ and $\t'=\t\cup\yv$.
  By (\ref{comp-arrow}), $v$ is computable. Since
  $\dom(\t)\sle\FV(u)\moins\FV(\vt)$ and $y\notin\FV(\vt)$,
  $\dom(\t')\sle\FV(yu)\moins\FV(\vt)$. Thus, by induction hypothesis,
  $(yu)\s\t'=vu\s\t'$ is computable. Since $y\notin\FV(u)$,
  $u\s\t'=u\s\t$. Thus, by (\ref{comp-red}), $u\s\t$ is
  computable.\qed

%\item [(subterm)] Since $\FV(u)\sle\FV(v)$,
%  $\dom(\t)\sle\FV(v)\moins\FV(\vt)$ and, by induction hypothesis,
%  $v\s\t$ is computable, hence strongly normalizable. Since
%  $\FV(u)\sle\FV(v)$, $u\s\t\tgt v\s\t$ and $u\s\t$ is strongly
%  normalizable. Since $u\s\t:B$ basic, by (\ref{comp-basic}), $u\s\t$
%  is computable.\qed
\end{lst}
\end{prf}

%%%%%%%%%%%%%%%%%%%%%%%%%%%%%%%%%%%%%%%%%%%%%%%%%%%%%%%%%%%%%%%%%%%%%%%%%%%%%%
% dependency pairs
%%%%%%%%%%%%%%%%%%%%%%%%%%%%%%%%%%%%%%%%%%%%%%%%%%%%%%%%%%%%%%%%%%%%%%%%%%%%%%

\section{Higher-order dependency pairs}

In the following, we assume given a set $\cR$ of admissible rules. The
sets $\fap(t)$ of {\em full application positions} of a term $t$ and
the {\em level} of a term $t$ are defined as follows:

\begin{lst}{--}
\item $\fap(x)=\vide$ and $\level(x)=0$
\item $\fap(\lx t)=1\cdot\fap(t)$ and $\level(\lx t)=\level(t)$
\end{lst}

\noindent If $f\in\cD$ then:
\begin{lst}{--}
\item $\level(ft_1\ldots t_n)=1+max\{\level(t_i)~|~1\le i\le n\}$
\item $\fap(ft_1\ldots t_n)=\{\vep\}\cup\bigcup_{i=1}^n1^{n-i}2\cdot\fap(t_i)$
\end{lst}

\noindent If $t\neq ft_1\ldots t_n$ with $f\in\cD$, then
$\fap(tu)=1\cdot\fap(t)\cup 2\cdot\fap(u)$ and
$\level(tu)=max\{\level(t),\level(u)\}$.

\begin{dfn}[Dependency pairs]
The set of {\em dependency pairs} is $\DP= \{l\a r|_p~|~l\a
r\in\cR,\, p\in\fap(r)\}$. The {\em chain relation} is
$\ach=\ari^*\adph$, where $\ari$ is the restriction of $\ar$ to
non-top positions, and $\adph$ is the restriction of $\adp$ to top
positions.
\end{dfn}

If, for all $l\a r\in\DP$, $\FV(r)\sle\FV(l)$, we have
${\ach}\sle{\ar^+\tge}$. Hence, $\abch$ is terminating whenever $\abr$
so is. We now prove the converse:

%%%%%%%%%%%%%%%%%%%%%%%%%%%%%%%%%%%%%%%%%%%%%%%%%%%%%%%%%%%%%%%%%%%%%%%%%%%%%%
% dp theorem
%%%%%%%%%%%%%%%%%%%%%%%%%%%%%%%%%%%%%%%%%%%%%%%%%%%%%%%%%%%%%%%%%%%%%%%%%%%%%%

\begin{thm}
\label{thm-dp}
Assume that, for all $l\a r\in\cR$ and $p\in\fap(r)$,
$\FV(r|_p)\sle\FV(r)$ and $r|_p$ has the type of $l$ (*). Then, $\abr$
is terminating if $\abch$ so is.
\end{thm}

\begin{prf}
By (\ref{comp-sn}), this is so if every term is computable wrt
$\ar$. By (\ref{comp-all}), this is so if every $f^{\vT\A B}\in\cD$ is
computable. By (\ref{comp-arrow}), this is so if, for all $\vt:\vT$
computable, $f\vt$ is computable. We prove it by induction on
$(f\vt,\vt)$ with $(\ach,(\abr)\lex)\lex$ as well-founded ordering
(H1). Indeed, by (\ref{comp-sn}), $\vt$ are strongly normalizable wrt
$\abr$. By (\ref{comp-neutr}), it suffices to prove that every reduct
of $f\vt$ is computable. If $\vt\abr\vt'$ then, by (H1), $f\vt'$ is
computable since, by (\ref{comp-red}), $\vt'$ are computable and
${\abch\!(f\vt')}={\abch\!(f\vt)}$. Now, assume that there is $f\vl\a
r\in\cR$ and $\s$ such that $\vt=\vl\s$. Since rules are admissible,
by Lemma \ref{lem-comp-pat}, $\s'=\s|_{\FV(r)}$ is computable. We now
prove that $r\s'$ is computable by induction on the level $n$ of $r$
(H2). Let $p_1,\ldots,p_k$ be the positions in $r$ of the subterms of
level $n-1$; $\vy^i$ be the variables of $\FV(r|_{p_i})\moins\FV(r)$;
$x_1,\ldots,x_k$ be distinct variables not occuring in $r$; $r'$ be
the term obtained by replacing $r|_{p_i}$ by $x_i\vy^i$ in $r$; and
$\t=\{x_i\to \la\vy^i r|_{p_i}\s'\}$. We have $\level(r')=0$ and
$r'\s'\t\ab^* r\s'$. If $\t$ is computable then, by
(\ref{comp-no-fun}), $r'\s'\t$ is computable and we are done. By (*),
$\{\vy^i\}=\vide$ and it suffices to prove that $r_{p_i}\s'$ is
computable. For all $i\le k$, $r|_{p_i}$ is of the form $g\vu$ with
$\level(u_j)<n$. By (H2), $\vu\s'$ are computable and, since $f\vt\ach
r|_{p_i}\s'$, by (H1), $x_i\t$ is computable.\qed\\
\end{prf}

The condition on free variables is an important restriction since it
is not satisfied by function calls with bound variables like in
$(lim\,F)+x\a lim\,\la n(Fn+x)$.

%%%%%%%%%%%%%%%%%%%%%%%%%%%%%%%%%%%%%%%%%%%%%%%%%%%%%%%%%%%%%%%%%%%%%%%%%%%%%%
% manna-ness
%%%%%%%%%%%%%%%%%%%%%%%%%%%%%%%%%%%%%%%%%%%%%%%%%%%%%%%%%%%%%%%%%%%%%%%%%%%%%%

\begin{thm}
\label{thm-manna}
An higher-order reduction pair is two relations $(>,\ge)$ such that:

\begin{lst}{--}
\item $>$ is well-founded and stable by substitution,
\item $\ge$ is a reflexive and transitive rewrite relation containing $\ab$,
\item ${\ge\circ>}\sle{>}$.
\end{lst}

In the conditions of Theorem \ref{thm-dp},
$\abch$ terminates if ${\cR}\sle{\ge}$ and ${\DP}\sle{>}$.
\end{thm}

\begin{prf}
By (\ref{comp-sn}), this is so if every term is computable wrt
$\ach$. By (\ref{comp-all}), this is so if every $f^{\vT\A B}\in\cD$
is computable. By (\ref{comp-arrow}), this is so if, for all $\vt:\vT$
computable, $f\vt$ is computable. We prove it by induction on
$(f\vt,\vt)$ with $(>,(\abr)\lex)\lex$ as well-founded ordering
(H1). Indeed, by (\ref{comp-sn}) and Theorem \ref{thm-dp}, $\vt$ are
strongly normalizable wrt $\abr$. By (\ref{comp-neutr}), it suffices
to prove that every reduct of $f\vt$ is computable. If $\vt\abr\vt'$
then, by (H1), $f\vt'$ is computable since, by (\ref{comp-red}),
$\vt'$ are computable and ${>\!(f\vt')}\sle{>\!(f\vt)}$ since
${\abr}\sle{\ge}$ and ${\ge\circ>}\sle{>}$. Now, assume that there is
$f\vl\a r\in\DP$ and $\s$ such that $\vt=\vl\s$. Since rules are
admissible, by Lemma \ref{lem-comp-pat}, $\s'=\s|_{\FV(r)}$ is
computable. Since ${\DP}\sle{>}$ and $>$ is stable by substitution,
$f\vt>r\s'$. Thus, by (H1), $r\s'$ is computable.\qed\\
\end{prf}

An example of reduction pair can be given by using the higher-order
recursive path ordering $>\horpo$ \cite{jouannaud99lics}. Take
$>=(\ab\cup>\horpo)^+$ and $\ge=(\ab\cup>\horpo)^*$. The study of
these two relations has to be done. However, $>\horpo$ does not take
advantage of the fact that $>$ does not need to be monotonic. Such a
relation is given by the weak higher-order recursive computability
ordering $>\whorco$, whose monotonic closure strictly contains
$>\horpo$ \cite{blanqui06horpo}. Moreover, $>\whorco$ is transitive,
which is not the case of $>\horpo$. It would therefore be interesting
to look for reduction pairs built from $>\whorco$.

%%%%%%%%%%%%%%%%%%%%%%%%%%%%%%%%%%%%%%%%%%%%%%%%%%%%%%%%%%%%%%%%%%%%%%%%%%%%%%
% biblio
%%%%%%%%%%%%%%%%%%%%%%%%%%%%%%%%%%%%%%%%%%%%%%%%%%%%%%%%%%%%%%%%%%%%%%%%%%%%%%

%\bibliographystyle{plain}
%\bibliography{abbrev,short-names,mybib}

\end{document}